# Plastic computing, the cloud continuum journey beyond infinity

Xavi Masip-Bruin[1], Jordi Garcia[1], Adrián Asensio[1], Francesco D'Andria[2], Admela Jukan[3], Shahrok Daijavad[4], Panos Trakadas[5]

[1] *Universitat Politècnica de Catalunya (UPC), CRAAX Lab, Vilanova i la Geltrú, 08800, Spain*
[2] *Eviden, BDS INN R&D (formerly called ATOS), Barcelona Spain*
[3] *Technische Universität Carolo-Wilhelmina zu Braunschweig, Germany*
[4] *IBM, Almaden Research Center, San Jose, USA*
[5] *National and Kapodistrian University of Athens, Department of Port Management and Shipping, Dirfies Messapies, Greece, 34400*

*Abstract* – The ever-increasing challenges introduced by the diversity of current and envisioned network technologies and IT infrastructure draw a highly distributed and heterogeneous topology where innovative services must be optimally deployed to guarantee maximum level of quality for users. Indeed, paradigms such as the cloud continuum, bringing together edge and cloud computing, along with the new opportunities coming out by considering non-terrestrial networks connecting future 6G ecosystems, all with no doubt facilitate the development of innovative services in many different areas and verticals. However, considering the intensive data and quality requirements demanded by these services, the distribution of the execution tasks must be optimally designed. On the infrastructure side, several initiatives are already active aimed at providing a Meta-OS that may seamlessly manage the different actors (services, infrastructure and users) playing under this paradigm. However, several aspects remain yet limited, particularly when referring to the "mapping" of resources into services, where innovative technologies based on bidirectional coordination and modeling may be pivotal for an optimal performance. In addition, the upcoming demands coming from the adoption of network technologies easing users' connection with high levels of quality, such as 6G, as well the study of NTN opens up the traditional cloud continuum to include also satellites that may extend the cloud paradigm further than ever considered. This paper shows a seed work toward an extendable paradigm so called as plastic computing whose main objective is to optimize service performance and thus users' satisfaction, through considering a bidirectional modeling strategy, easily extendable to adopt novel network and IT technologies and paradigms. Finally, two examples are briefly introduced to highlight the potential benefits of the plastic computing adoption.

*Keywords*- **IoT/Edge/Cloud, Cloud continuum, plasticity, correlated orchestration**

## I. Introduction: The Continuum Landscape

Today, the wide adoption of cloud infrastructures to support the computing and storage demands required by many different applications from many distinct sectors is undeniable. In fact, there is no reasonable doubt on the large benefits cloud computing brings to users willing to optimally and efficiently run applications, what is actually demonstrated by the large impact cloud computing related business have on the global economy [1]. However, although broadly recognizing these benefits, some specific needs and demands, mainly related to network consumption, data freshness, data privacy, applications latency and security concerns, are continuously demanding for a different or enhanced computing paradigm. Indeed, a different, top-down shift on the traditional cloud computing architecture definition, poses edge computing as the complementary computing paradigm to locally manage data processing, reducing latency, and limiting the network load while also contributing to improve the privacy and security guarantees. Thus, two computing paradigms, cloud computing and edge computing, along with the diversity, heterogeneity and the usual mobility characterizing the IoT devices, should live together, becoming the key ingredients of the paradigm defined as cloud continuum (also known as cloud-edge, IoT/edge/cloud or fog-to-cloud–F2C), and hence requiring for novel control and management solutions to properly handle these ingredients. Efforts to be invested on designing such novel control and management strategies must also accommodate the continuously evolving continuum scenario, mainly represented by three components, the network evolution, the ever more resources demanding services and data-intensive applications, as well as the underlying technologies to be used.

From the network perspective three key elements in the path toward softwarization must be highlighted. First, network disaggregation [2] is usually referred to as the separation of hardware and software components, enabling a modular network setup where network functions and services can run independently across different physical and virtual assets. Among others, the disaggregation concept, becomes key for the development of the future optical networks [3], datacenters [4] and 6G ecosystems [5]. However, network disaggregation stresses the need to manage a highly distributed and virtual landscape, introducing new challenges, such as ensuring interoperability, maintaining consistent performance levels, and securing a more complex network architecture that could be vulnerable to novel cyber threats. Second, the network slicing concept enables service providers to flexibly offer customized networks with different functionalities for either supporting diverse services or serving groups of users with specific service requirements [6]. However, the fact that common infrastructure and resources are shared amongst network slices needs to be carefully studied in order to, for example, guarantee network slices to be dynamically created, monitored and safeguarded. Finally, the highly distributed nature of the current network, with different technologies, domains and segments, along with the trend toward networks convergence, considering terrestrial networks and satellite networks (NTN), depicts an scenario where the current cloud layout may be opened up to include satellites (in different orbits), defining a larger continuum, what with no doubt, poses the need for network functions to be composed flexibly and dynamically, based on service needs in diverse cloud and edge environments [7].

Simultaneously, the non-stoppable advent of new applications and services into the market are, with no doubt, stressing the management of those resources needed to accommodate strict demands. Indeed, paradigms, such as the metaverse [8], are opening an enormous door to an unforeseen set of applications, such as X-AR, requiring ultra-real-time reaction (demanding for non-perceptible latency and high computational processing), virtual entities mobility management (i.e., holograms), extreme data management (including data spaces), or extremely dynamic set of virtual resources deployment, sharing and interoperability.

Moreover, the deployment of a diverse set of technologies aimed at improving services efficiency and performance, are also challenging current ICT infrastructures to support their specific needs. For example, the widely accepted utilization of AI brings in specific demands to allocate the computational load to train models and the proper resources to store and preserve data, while guaranteeing data privacy and freshness. Strategies, such as Federated Learning, (FL), Reinforcement Learning (RL), Active Learning (AL) and Continuous Learning may make the most out of the cloud continuum resources to this end. Linked to AI, Digital Twin appears as a sister technology, useful for systems modelling, extremely relevant in the development of proactive solutions (e.g., predictive maintenance, early attacks detection, performance estimations, etc.). Another innovative technology example sits on the trend toward services containerization that, although it is aimed at facilitating a transparent deployment of services, it is also imposing governance policies on how the different solutions for containerizing services may interact each other and on how available resources are clustered to be efficiently managed.

Finally, recognized the challenging scenario posed by the cloud continuum landscape, from a research perspective, the final blow comes up when considering security aspects, which raise unprecedented doubts, e.g., how the security issues in the different technologies deployed in the distinct cloud continuum infrastructure segments are correlated? how security is managed and certified in AI systems?, or how constrained edge devices handle security needs to avoid becoming bots supporting large attacks?, just to name a few that must be, with no doubt, carefully analyzed.

Therefore, it is clear that the novel contextual flavours added to the cloud continuum will shape the approach to maximize the cloud continuum benefits towards a higher shift, mainly challenged by: *i*) the dynamicity of physical and virtual resources; *ii*) the need for a dynamic and ad-hoc resources and services shaping under a common approach, accommodating existing and unforeseen policies and rules (e.g., security, green); *iii*) technology transparency, and; *iv*) a robust approach where any action aimed at disturbing business continuity is either prevented or in the worst case mitigated. This novel scenario is conceptualized under an extended computing paradigm that while being extremely ductile becomes very robust. In this paper, we propose to name this paradigm as Plastic Computing. In short, Plastic Computing refers to the heterogeneous, volatile and highly dynamic computing paradigm that results from bringing together all resources within the continuum (both infrastructure - real and virtual - and data, all distributed from the IoT layer to the cloud), managed through a set of specific ad-hoc and bidirectionally designed, composed, orchestrated and deployed functionalities open to meet a set of specific policies, to optimally execute a set of partially yet unforeseen innovative services.

Indeed, the envisioned scenario requires main players in the continuum, namely resources, management functions and services, to be all orchestrated under a continuous and correlated approach. Plastic Computing proposes to build such a novel approach through a loop, implementing the mutual and bidirectional concepts in the decision-making process for the continuum players. First, resources should be endowed with the ability to be continuously shaped and moulded to support the specific needs of highly demanding services, both during the allocation and runtime windows. Second, management functions should be dynamically formed and deployed to meet specific resources and services demands. Finally, services should be intelligently partitioned into jobs and deployed through virtual components (i.e., containers) to optimise their execution (performance, energy efficiency, etc.).

To this end, this paper presents the need for such an evolved, dynamic and adaptable computing paradigm, introducing a functional architecture for the Plastic Computing concept, including a preliminary set of functional blocks designed to facilitate an adaptable, bidirectional management and orchestration, enriched with a preliminary emulated testing environment to facilitate proactivity.

The rest of the paper is organized as follows. Section II gives an overview of the envisioned context, identifying new expected needs ending up in a set of challenges. Section III proposes a functional architecture for the proposed plastic computing concept, including a few key references to the related work, along with an illustrative example of a coordinated control strategy. Section IV shows preliminary conceptual results obtained by considering the deployment and execution of two applications, showing the benefits of the proposed paradigm. Finally, Section V concludes the paper.

II. THE ENVISIONED CONTEXT: REQUIREMENTS, NEEDS AND LIMITATIONS

As introduced in the previous section, the ever-unstoppable evolution in the seed foundations of IT systems (easing for example ubiquitous high quality communication through the 5G/B5G/6G paradigms, or the large deployment of far edge and IoT systems endowed with high computational capabilities supported by recent advances in chips design), along with a new era where new services are enriched with technological advances such as AI, are with no doubt setting the ground for a novel computational paradigm featured by extremely demanding execution requirements. From a technical perspective, we describe this paradigm as plastic, considering

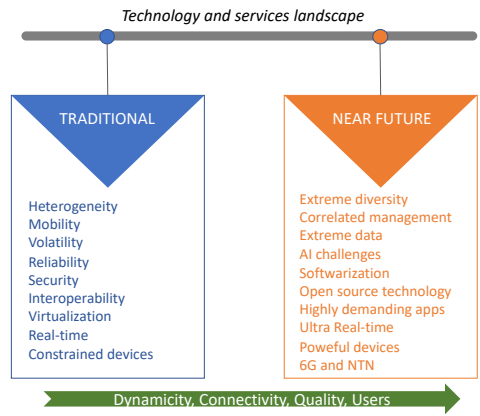

**Fig. 1.** The Cloud continuum landscape

keywords, such as dynamic, heterogeneous, bidirectional, adaptive, intelligent, efficient, optimal, modular, reliable, secure, robust and virtual, as the key ingredients toward the Plastic Computing concept. The challenges linked to this plasticity sit on four key principles, to be later mapped into a functional architecture for plastic computing: *i*) AI-assisted proactiveness (e.g., optimal decision-making strategy, efficiency); *ii*) forecasting actions (e.g., anticipated detection for execution anomalies, security risks, or reinforcement for QoS/SLA support); *iii*) extremely high virtualization (e.g., portability, logical correlation, interdependency); and *iv*) runtime analytics (e.g., sandbox environment, instant monitoring, deep observability).

Many efforts have been done in terms of scientific papers and research initiatives addressing the need for a complete and innovative cloud continuum management solution. Beyond the large track of papers that may be found in the literature addressing specific challenges in this area, large research projects, such as several EU funded projects, are aimed at designing and developing a so-called Meta-OS, thus extending the traditional control and management plane to build an innovative high-level OS. The ICOS project [9] extends preliminary efforts toward designing a control and management plane for the cloud continuum environment [10], by considering the fact that the existing resources building the cloud continuum stack are usually managed by other native OSs, so the real need is to identify high level orchestration functions that may be deployed on top of these OSs in a coordinated way. The FLUIDOS project [11], proposes a definition of the continuum that focuses on three properties, namely orchestration, communication, and transparency, borrowing from the seminal concept of liquid computing described in [12]. The NEMO project [13], assumes the orchestration of services into clusters of nodes (even when dispersed across different locations and infrastructure providers), and in order to ensure efficient and high-quality service delivery, works on the design of a Meta-OS, which will enable multi-cluster and multi-network orchestration of containerized workloads across the cloud continuum.

However, beyond recent efforts focusing on current challenges in the cloud continuum arena as demanded by services and applications developed so far, the envisioned next generation of services are expected to be even more demanding in terms of optimization, dynamicity, human-interaction, adaptability, and security, among others; therefore limiting the impact current solutions may have to address these challenges. Moreover, this added complexity is even exacerbated when considering the specific needs imposed to the overall computational infrastructure, driven by the adoption of innovative technologies, such as the risks associated to the utilization of AI, or the mandatory inclusion of security, privacy and reliability, all becoming fundamentals pillars in the development of any computational infrastructure management solution. Figure 1 shows the so-called traditional attributes linked to the cloud continuum paradigm as the main driver for the ongoing research initiatives, as well as the so-called near future challenges, responsible for fueling additional research in the area. A clear link among them may be easily observed. For example, while current efforts address devices heterogeneity and mobility considering constrained devices at the edge, the future landscape envisions an extremely diverse scenario, including powerful devices at any layer of the continuum stack. Moreover, widely accepted trends in the ICT arena, such as softwarization, disaggregation and open source, are imposing specific needs and thus, posing clear challenges in systems virtualization, interoperability and operation. Specifically, the envisioned challenges may be summarized as follows:

- Need for continuous proactivity to facilitate an optimal and accurate multi-dimensional decision process. This assessment refers to addressing several specific challenges, such as how to leverage RL to ease the interaction with the multi-dimensional context, how to leverage AL to minimize the data to be used for training and how these data may be labelled, or how to efficiently allocate computational processes in distributed learning to facilitate data privacy while benefitting from powerful enough edge systems.
- Need for a pre-tested decision-making process to maximize the accuracy of any decision-making process, addressing challenges related to dynamic and continuous mirroring of the cloud continuum stack while simultaneously monitoring services performance.

- Need for a continuous pre-real-time adaptation of resources and services in a correlated way, also addressing the challenges linked to added-value attributes and policies resources will be characterized with (e.g., green, etc.), and the innovative services to come, with capacity to continuously map resources status into services management, under a bidirectional and ad-hoc paradigm.
- Need for developing a human-in-the-loop intelligent orchestration system, supported by AI-based functionalities, facilitating human awareness, and thus paving the way for different levels of human control depending, among others, on how critical or sensitive the ICT infrastructure and the distinct services to be managed are.
- Need for an expandable, modular, adaptable, open and easy to interface solution, able to accommodate existing and even non-envisioned requirements. To this end, challenges related to identifying a boundless set of ad-hoc management primitives that enriched with a composer function may deal with the expected diversity and needs of the services to be executed.
- Need for continuous monitoring and benchmarking processes that may, on one hand, facilitate deep observability on telemetry and thus collecting and being able to handle the required data for further analytics, and, on the other hand, endorse decisions taken with guarantees about specific ad-hoc defined attributes, such as performance or cost.

To address the above challenges, we believe that a real need has emerged to manage the different cloud/edge/IoT/data resources as well as the highly demanding emerging services, under a novel framework capable of bidirectionally orchestrate and on-demand compose physical and virtual resources, and services, to meet current and future needs. Moreover, the envisioned management framework must also accommodate solutions to handle the challenges imposed by the adoption of new technologies, guaranteeing these technologies not to weak the overall performance. A key example of this assessment refers to the widely reported set of AI incidents related to AI robustness, for example legal hallucinations in Large Language Models (LLMs) that have been found to be alarmingly prevalent, occurring from 58% to 88%, depending on the model used [14], or the deployment of Adversarial Attacks [15] to interrupt systems and services continuity.

### III. DIVING INTO THE PLASTIC COMPUTING CONCEPT: FROM STATIC TO BIDIRECTIONAL COORDINATION

The scenario described above fuels the need for an even more dynamic paradigm. Indeed, unlike current static solutions on the cloud-edge arena, mainly driving toward the design of a coordinated management plane, currently extended with the Meta-OS trend, in this paper we propose a new strategy, where dynamicity is handled bidirectionally and thus, resources (the whole continuum stack) and services (jobs) are mutually and under a correlated context, dimensioned and orchestrated. Indeed, while the traditional research question is how resources may be managed to optimize services execution, nowadays question is how resources and services may be orchestrated and dimensioned, dynamically, to optimize services execution and resources management. In fact, this novel approach may not only impact on infrastructure and services performance parameters but may also push for new business models supported by potential users (individuals but also companies or the public sector) willing to share their infrastructure (ad-hoc created clusters) under a very dynamic, but controlled, management strategy.

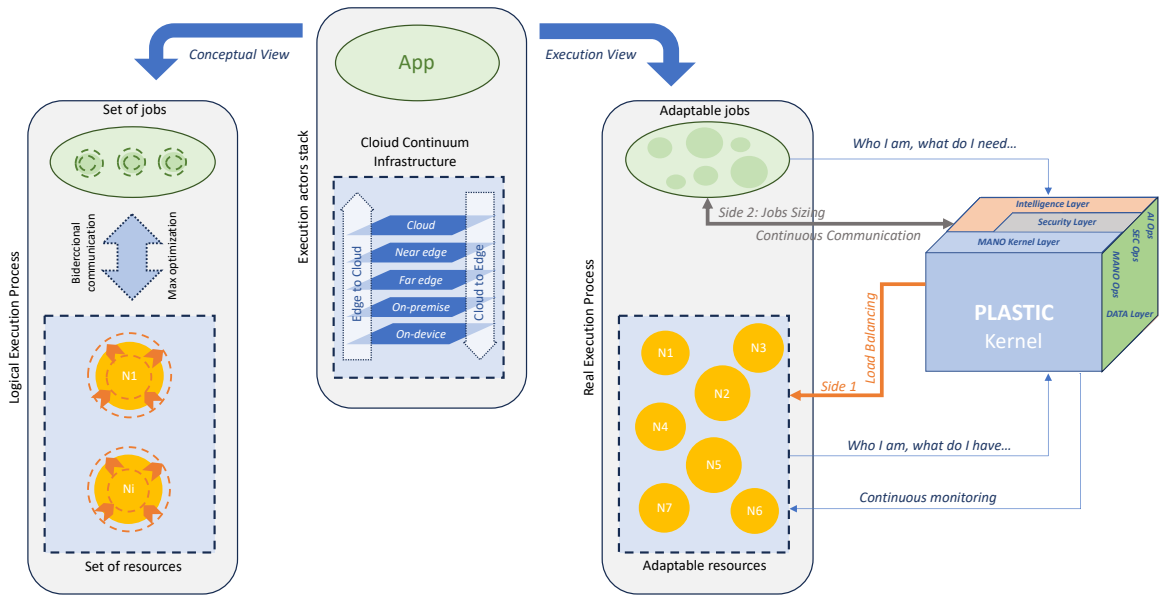

**Fig. 2.** The Plastic Computing Concept

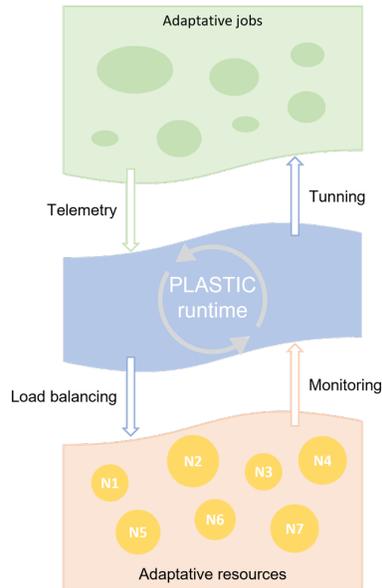

**Fig. 3.** The Bidirectional Coordination Concept

Figure 2 graphically depicts the concept behind the Plastic Computing idea. We start identifying the two key actors in the stack, namely the applications/services to be executed and the entire cloud continuum infrastructure. The figure introduces the plastic concept through both conceptual (logical) and real execution views. The conceptual view (shown on the left) introduces the dynamicity context to be deployed at both actors in a correlated and coordinated way in order to maximize the overall performance, graphically representing the capacity to make the different components (jobs and resources) to be adapted to the mutual needs. The real execution view (shown on the right), presents the two main concepts proposed in the Plastic Computing paradigm. The first, bidirectional coordination refers to the continuous and mutually correlated adaptation of the resources and services to maximize the overall performance. Figure 3 describes in more detail the bidirectional coordination loop, from resources to jobs, including interfacing functions. The second, corresponds to a novel orchestration component, identified as a Plastic Kernel, that includes all management and orchestration (MANO) functions, supported by both security and intelligence functionalities. The expected bidirectionality is described through two execution workflows represented as two sides (directions), for load balancing and jobs sizing respectively. The overall process will start setting a continuous communication between the kernel and the application to be executed as well as a continuous monitoring strategy between the kernel and the set of resources. While many existing technologies may be deployed, extended or modified, to support the need for such a monitoring process (many of them depicted in Figure 4), a solution must be designed to implement the continuous communication between the application and the kernel. This communication must facilitate the kernel to decide how "large" the jobs and how "powerful" the resources should be to accommodate all services needs, resources availability as well as the proper load balancing policies toward an optimal performance. A fundamental ingredient in these policies is the strategy to define the bidirectionality, i.e., how the shaping of resources and jobs is jointly and correlatedly decided. To this end, we propose to include a Matrix Fabric into the Plastic Kernel (see Figure 4) that will bring together the set of mechanisms, strategies, learning processes and policies to define such envisioned mutual shaping.

In summary, we may highlight several key challenges that must be addressed to facilitate a wide and optimal development of the key Plastic Computing concept:

- Identify a low-computational-cost strategy for guaranteeing the continuous communication between the application and the Plastic Kernel
- Define the multimodal correlation functionality that may support the mutual adaptation and shaping of resources and jobs, transparent to any kind of service and resource, defined as bidirectional coordination
- Define a proactive layout, based on a pre-testing idea, supported by an emulated environment mirroring the real context, easing an accurate, optimal and safe decision-making process while simultaneously guaranteeing ultra-real-time processing
- Define the set of robustness policies that must make the overall computing context to be safe, secure and reliable
- Define proper strategies to make the most out of the large set of resources for models training, particularly leveraging innovative approaches, such as those based on Federated Learning, Active learning, Reinforcement Learning or even Continual Learning.

*A) Key functional blocks in the conceptual architecture*

In this paper we aim at introducing the key expected challenges driving the need for a conceptually much more ambitious strategy for services and continuum resources orchestration. As discussed in the previous section, this need is motivated by the even non yet foreseen requirements posed by the new expected services. In this context, we identify plasticity as the main concept to be considered to suitably adopt the expected characteristics, such as an extremely accurate mutual correlated decision-making process and an ultra-real-time decision process supported by pre-testing analysis.

To support these needs, this subsection introduces key functional blocks of a proposed architecture, putting together the high-level functionalities we consider as a "must have" for any potential deployment. Figure 4 depicts such proposed architecture, including the whole set of expected functionalities to accommodate the envisioned challenges, interfaces the services on the North and the infrastructure on the South. The proposed architecture is sector agnostic, hence no matter what the domain is, the distinct functionalities should be learnt to meet their specific needs and requirements. On the other hand, the infrastructure should encompass all potential devices composing the continuum, from the IoT up to the cloud. In fact, the figure includes a potential distribution and classification of a candidate set of devices into different categories, identifying main characteristics and proposing some examples on each of them. The Plastic Kernel is functionally split into two main clusters, one oriented to provide the real execution context functionalities and the other one to support the pre-testing functionalities though the definition of an emulated context. The cluster of components identified as Execution Context in the figure puts together three main modules. The one on the top in the figure, responsible for interfacing the services to be executed, deals with all technologies and tools considered for services deployment, what includes tools for containerizing services (e.g., Dockers), tools for managing containers and clusters of containers (e.g., Kubernetes) as well as tools for specific management of cloud continuum distributed services (e.g., NuvlaBox). All these tools and those to come up must be seamlessly integrated into the Plastic architecture. The module on the bottom of the figure, identified as Resource Management, must consider all tools easing the interface with the infrastructure while also providing the data required by the Plastic Computing Kernel to manage it, including for example tools for monitoring such as Prometheus. Finally, the third module, identified as the Mutual Orchestration Engine (MOE), is the responsible for developing the mutual and bidirectional services to resources mapping. To this end, a Matrix Fabric is proposed as the key component responsible for delivering the ad-hoc, correlated and adaptable mapping. This Matrix Fabric will interact with the different functionalities designed to manage the overall system. In order to support the highly dynamicity to be supported by the complete system, we envision a dynamic approach, where the proper orchestration functionality to be applied properly matching the specific and real-time context conditions, will be composed leveraging a set of so-called orchestration primitives (including seed functions such as, slicing, offloading, models, clustering, etc.). This flexibility is proposed as a solution to adapt the proposed solution to the envisioned highly changing scenario, not only in terms of resources availability but also in terms of jobs needs (e.g., SLA or security compliance), and even users demands (e.g., expected QoE/QoS). The MOE module also includes three additional components. The Programming model will deal with the specific functionalities linked to runtime execution, thus is close interaction with the Matrix Fabric. The AI Box and Security Box, will take over the specific needs related to AIOps and security provisioning, thus also impacting the Matrix Fabric to guarantee proper AI processes and a secure scenario.

As said above, the Execution Context cluster of components deals with the real services execution. However, since a proactive strategy is considered as a must to guarantee an optimal services execution, and considering the ultra-real-time needs expected in new services to come up, the Emulated Context is proposed as the right spot where all pre-testing functionalities must be deployed. To this end, a Sandbox scenario is proposed where a properly mirrored copy of the overall real context is created. Technologies supporting Digital Twins are key to facilitate this proactive process. In this context, two main functionalities are expected. The first is responsible for continuously monitoring the performance of the emulated context, so that any deviation may be early detected. The second is responsible for testing the impact any proposed action may have before being deployed on the real layout, leveraging specific processes based on benchmarking and impact analysis.

These two clusters of components leverage two key conditions. The first deals with continuous learning and feedback process set between the emulated and the execution context. This process is mandatory for a successful deployment of the solutions proposed to address the specific Plastic Computing challenges. This learning will affect all functionalities to be composed. The second refers to how data is managed, so the proper data governance strategies may be designed to accommodate the specific needs Plastic Computing brings in.

*B) Briefing the Expected Benefits*

A correct deployment of the proposed plastic computing paradigm should enforce a mutual characterization of both available resources and services to be executed, aimed at optimizing the deployment of services into the resources, using only those services that may maximize services performance. To this end, the current paradigm where resources are characterized and selected to optimally allocate services needs, should be extended to consider a bidirectional coordination, that will require the setting of different models, inferred from past performance patterns, based on deploying AI-assisted analytics.

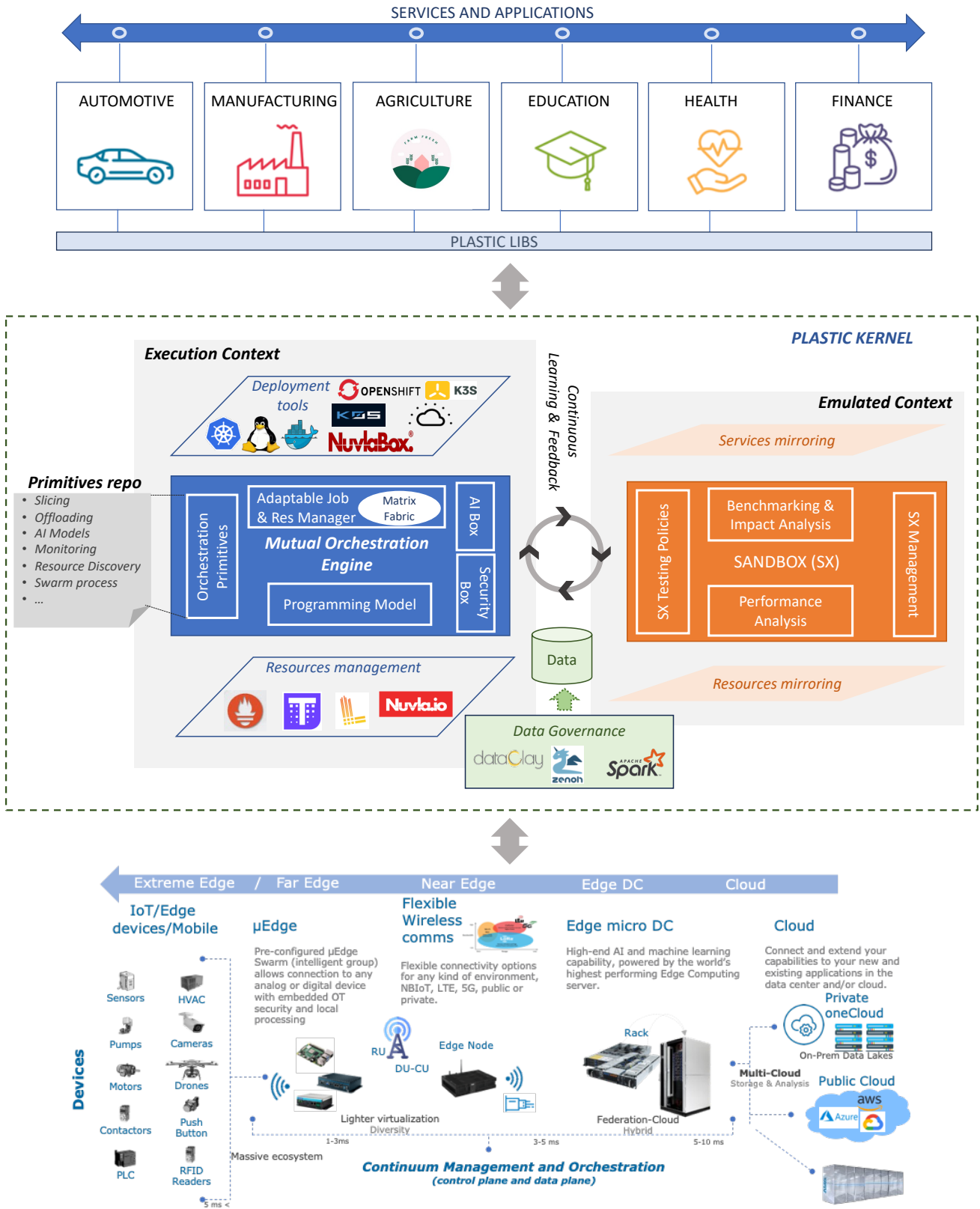

**Fig. 4.** Functional architecture for the Plastic Computing scenario

## IV. PERFORMANCE ANALYSIS: ILLUSTRATIVE EXAMPLE

As introduced in the previous sections, the concept of Plastic Computing defines an optimization cycle where resources are allocated for optimal application execution, while applications are configured for optimal resource utilization. Two different experiments will be used to illustrate this idea as a proof of concept.

The first experiment analyzes how resources allocation can determine the execution performance of applications. In this experiment, a parallel version of a Fast Fourier Transform (FFT) algorithm is computed over a three-dimensional dataset with 256 signals per dimension. This code has been executed on a parallel cluster with 12 nodes. Fig. 5.a shows the execution time (in seconds) of FFT for different executions, ranging from 2 to 12 cores. As it can be seen in the figure, the best performance has been obtained by running the application using 5 cores. With fewer cores, the execution is under powered because it has more potential for parallel execution; however, with more cores, the overhead of parallel coordination becomes too high due to the size limitation of this dataset. Consequently, adding more resources does not always derive in an enhanced performance. Note that this result is specific to this particular application and data size instance; the same application with a different data size would expose a different degree of parallelism.

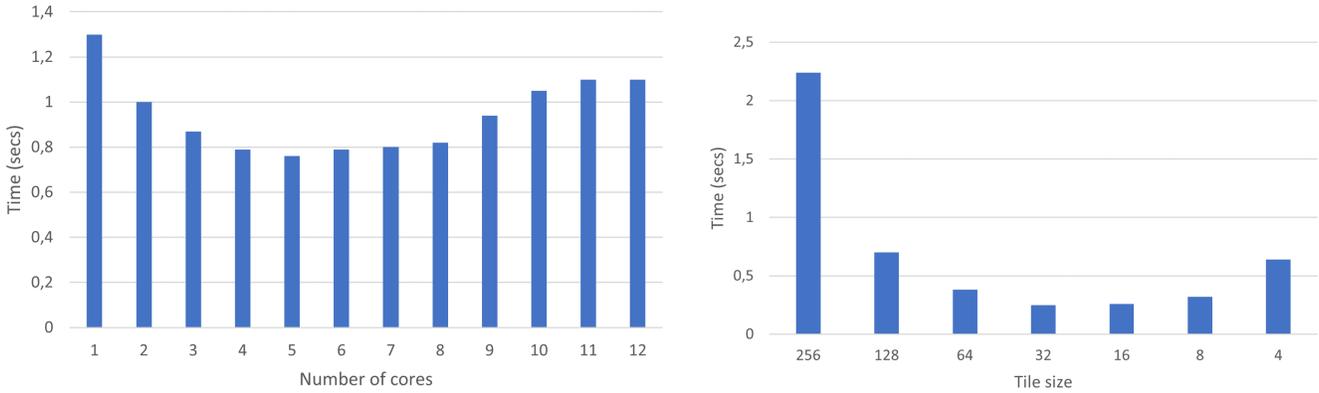

**Fig. 5**. a) Execution time of FFT with different number of cores; b) Execution time of MS with different tile sizes.

The second experiment illustrates how proper application tuning can also impact execution performance for a given resource topology. In this experiment, a two-dimensional Mandelbrot Set (MS) computation has been used running on a 1024x1024 dataset. The MS computation is an embarrassingly parallel iterative task decomposition problem, and different task sizes (this is, the tile size to be assigned to each node) can be defined to determines the granularity of each task. This code has been executed on a parallel cluster with 8 nodes. Fig. 5.b shows the execution time (in seconds) of MS for different tile sizes, ranging from 256 (very small number of large tiles) to 4 (very large number of small tiles), reducing the tile size in each execution by half. As can be observed in the figure, the optimal performance has been achieved for 32-point tiles. With larger tiles (smaller number of tiles) the parallelism cannot be fully exploited; and with smaller tiles (larger number of tiles and, therefore, more tasks to be created) the overhead of task management absorbs the benefits of a higher degree of parallelism.

As a summary, these experimental results aim to be a proof of concept to validate the Plastic Computing concept, this is, demonstrate that selecting the right amount of resources for a given application improves performance, as well as demonstrate that applying the right problem configuration for a given resource topology also contributes to a performance improvement.

## V. CONCLUSION

This paper introduces the plastic computing concept as an extension of the traditional cloud continuum paradigm, setting the foundations for a bidirectional coordination approach where services are tailored to available resources and vice-versa. A preliminary functional architecture is presented, describing the key modules plastic computing is expected to reside on. For the sake of illustration, two easy examples of services execution are included showing that the simple fact of increasing resources may not be the right strategy to optimize service execution and also that designing the proper service configuration tailored to the resource availability may substantially contribute to improve the overall performance. The paper is considered as a seminal work introducing the bidirectional coordination approach under the name of plastic computing. Current work goes on identifying and analyzing many challenges and extensions to move forward, particularly focusing on both enlarging the continuum to include the NTN ecosystem, and potential new services with highly demanding requirements.

## Acknowledgments

This work has been partially by the HE ICOS project, funded by European Commission with Grant Number 101070177, and for UPC authors by the Spanish Ministry of Science and Innovation under grant PID2021-124463OB-I00 funded by MCIN/AEI/10.13039/501100011033 and by ERDF A way of making Europe, as well as by the Catalan Government under contract 2021 641 SGR 00326.